# Cylindrical Metasurface for Efficient Traveling-wave Magnetic Resonance Imaging at 7 T

Kristina I. Popova, Georgiy A. Solomakha, Zicheng Wen, Mikhail M. Popov,
Xiaotong Zhang, *Senior Member, IEEE*, Stanislav B. Glybovski, and Yang Gao, *Member, IEEE*

***Abstract*—This research focuses on the design and evaluation of an ultrathin cylindrical metasurface for improving the transmit efficiency of traveling-wave magnetic resonance imaging (MRI) of the human brain. To improve efficiency, we matched a travelling waveguide mode to an electrically large, lossy dielectric load using a thin cylindrical metasurface, which occurs to be a task closely related to impedance matching in waveguide circuits in the microwave. This metasurface was designed as a compact and lightweight replacement for a high-permittivity dielectric waveguide previously proposed for the same purpose. The dispersion analysis showed that both structures (waveguide and metasurface) support a similar type of slow-wave propagation, characterized by a uniform magnetic field profile close to the cylinder axis. At the Larmor frequency, the longitudinal wavenumbers showed close agreement. Based on the optimized unit cell geometry of the periodic copper strip grid loaded with PCB capacitors, full numerical model of the cylindrical metasurface in the presence of a voxel human body model was constructed. We also compared the proposed metasurface with the dielectric waveguide in the traveling-wave setup experimentally, including *in vivo* measurements performed on a healthy volunteer. The proposed metasurface showed improved $B_1^+$ homogeneity (by 17.3%), transmit efficiency (by 27.4%), and SAR-efficiency (by 23%) compared to the dielectric waveguide. The proposed cylindrical metasurface, optimized for field enhancement in the human brain at 7 T in the traveling-wave excitation regime, can further improve the transmit efficiency and homogeneity in the region of interest compared to state-of-the-art structures for traveling-wave MRI, at the same time, granting the advantages of light weight and compactness. The proposed metasurface can be used for increasing the transmit efficiency of single-channel traveling-wave excitation setups suitable for prospective research and clinical MRI tasks.**

## I. INTRODUCTION

Ultra-high magnetic resonance imaging (MRI) is a rapidly developing technology which provides unpreceded accuracy of non-invasive diagnostics of the human body and brain tissues [1].

At ultra-high magnetic fields (7 T and above, corresponding to frequencies of 297.2 MHz and higher), radiofrequency (RF) field inhomogeneity becomes a major challenge due to a shortened wavelength in biological tissues [2]. This effect manifests itself in pronounced standing-wave patterns of the $B_1^+$ field caused by multiple reflections and interference within the human body and the RF shield, and therefore, spatially non-uniform excitation in body and brain MRI [3]. One of the methods for improving homogeneity is associated with design of transmit (Tx) coils.

The most accepted approach in the literature consists of using multi-fed array coils, which support $B_1^+$ field homogenization through optimal phase distribution of driving voltages (active RF shimming) [1], [4]. Despite wide optimization possibilities provided, multi-channel arrays are yet not widely used for clinical applications due to hardly predictable appearance of local electric field hotspots causing high specific absorption rate (SAR) [5] and difficulties to concentrate the RF field in electrically large objects [4]. Another reason for that is the need for engineering maintenance of a multichannel excitation system [1]. In contrast, single-channel excitation setups for 7 T MRI are in wide clinical use and provide attractive possibilities for improving medical diagnostics of brain diseases. Therefore, finding new methods for passive RF shimming, in which dielectric pads [6], [7], [8], high-permittivity elements [9], [10], metasurfaces [8], [11], [12] and passive resonators [13], [14], [15] are combined with a single-fed Tx antenna is an important research direction.

One of promising passive RF shimming strategies is called traveling-wave (TW) magnetic resonance imaging proposed in [3] in 2009. In the classic TW MRI, the inner bore (RF shield) of an MRI scanner performs as a metallic circular waveguide supporting the fundamental $TE_{11}$ mode to create RF excitation in the region of interest (ROI) at the Larmor frequency using typically only one transmit channel. In this configuration, the radiating antenna [16] launching the waveguide mode is located at the service end of the MRI scanner, away from the patient.

This work was supported in the numerical part by the Ministry of Science and Higher Education of the Russian Federation (Project FSER-2025-0009), in experimental part by National Natural Science Foundation of China (52307256). (Corresponding author: Stanislav B. Glybovski.)

Kristina I. Popova is with School of Physics and Engineering, ITMO University, St. Petersburg, Russia (e-mail: kristina.shin@metalab.ifmo.ru).

Georgiy A. Solomakha is with High-Field MR Center, Max Planck Institute for Biological Cybernetics, Tubingen, Germany (e-mail: georgiy.solomakha@tuebingen.mp.de).

Zicheng Wen is with Hangzhou Institute of Technology, Xidian University, Hangzhou, China (e-mail: zichengwen@stu.xidian.edu.cn).

Mikhail M. Popov is with School of Physics and Engineering, ITMO University, St. Petersburg, Russia (e-mail: mikhail.popov@metalab.ifmo.ru).

Xiaotong Zhang is with College of Electrical Engineering, Zhejiang University, Hangzhou, China (e-mail: zhangxiaotong@zju.edu.cn).

Stanislav B. Glybovski is with School of Physics and Engineering, ITMO University, St. Petersburg, Russia (correspondence e-mail: stas@itmo.ru).

Yang Gao is with Hangzhou Institute of Technology, Xidian University, Hangzhou, China (e-mail: gaoyang01@xidian.edu.cn).

Typically, patch [1] or dipole [17] antennas are used to launch the waveguide mode, which is further directed towards the ROI in the human body thanks to propagation in the metallic circular waveguide. In contrast to using local Tx resonant coils, which are known to create standing-wave patterns of electric current and near fields, TW MRI mitigates field inhomogeneity by relying on propagating electromagnetic fields (traveling waves), thereby offering improved RF field homogeneity at high Larmor frequencies.

An additional advantage of TW MRI is its non-local excitation scheme. Since the Tx antenna is located far from the center of the bore, this approach improves the subject's comfort and provides valuable space near the ROI for Rx-only RF coils, local $B_0$ shimming arrays or visual stimulus equipment. However, the Tx-efficiency of the classic TW MRI implementation appears to be sub-optimal due to the poor matching between the wave impedances of the waveguide mode and waves propagating in absorbing tissues (of the human body or brain) [5]. As a result, the signal-to-noise ratio (SNR) of obtained images using TW MRI is low compared to the excitation using local RF coils due to insufficient flip angles in the ROI.

Several approaches have been proposed to address this limitation. One strategy involves the use of coaxial waveguides positioned between the Tx antenna and the subject [18], [19]. These structures enhance power transmission into the ROI and increase the achievable RF magnetic field amplitude. However, this approach exhibits important drawbacks for human brain imaging at 7 T. In particular, strong reflections from the volunteers' shoulders generate longitudinal standing waves, degrading the excitation homogeneity. Moreover, the mechanical and geometrical complexity of coaxial waveguide configurations limits their use for routine clinical scans. Another method to enhance TW MRI efficiency employs passive resonant structures placed inside the bore of the MRI scanner near the ROI [13], [14]. Local passive RF arrays based on loop resonators [14] can concentrate RF energy within the array volume, thereby reducing sensitivity to reflections from distant anatomical structures such as the shoulders. Nevertheless, these systems introduce additional structural complexity and impose stringent requirements on the experimental setup. Dielectric materials with tailored geometries and high permittivity can be used as an alternative type of passive structures for improving RF field magnitude and homogeneity [6],[7],[20]. Dielectric materials placed near the human body or the head significantly reduce the standing wave effects in the RF field distribution. In [7], it was proposed to use a dielectric insert with high permittivity, located inside an MR-scanner bore, to improve Tx-efficiency and homogeneity of the RF field in the human brain at 7 T. The dielectric insert increases the penetration of traveling-wave power into the ROI, which leads to an increase in the magnitude of the RF magnetic field in excitation mode compared to the case without passive structures. Recently, another configuration of the dielectric structure has been introduced, which makes it possible to increase the efficiency of RF excitation in the traveling-wave mode for the study of the human brain at 7 T [20]. The work proposes the use of a local dielectric waveguide (DW) placed around the human head. This DW enhances RF field magnitude and homogeneity within the brain by exciting a hybrid $HE_{11}$ mode through mode conversion from $TE_{11}$ to $TM_{11}$ within the metallic cylindrical waveguide of the MR scanner. This approach substantially improves power transfer to the brain without increasing SAR.

Despite improving Tx-efficiency and homogeneity compared to classical TW-setup, dielectric structures for passive shimming suffer from several limitations. High-permittivity dielectric inserts are often bulky, occupy significant space inside the scanner bore, and may reduce the subject's comfort. In addition, water-based dielectric materials can introduce imaging artifacts and complicate practical deployment, thereby limiting their suitability for routine clinical applications.

Artificial dielectric materials and metamaterials with precisely engineered subwavelength unit cells, have been proposed to overcome the limitations of water-based dielectrics and for the strong control over electromagnetic fields [21], [22]. Their applications have progressed from the reactive near-field regime, where they are mainly used to reshape and enhance local RF fields [21], [23], [24], [25], to the radiative regime, where wave propagation can be manipulated more directly [26]. In particular, recent studies have demonstrated the feasibility of reproducing the function of dielectric waveguides using artificial materials, enabling control of propagating waves inside the MR bore-waveguide [26]. Although early implementations often relied on bulky three-dimensional structures [23], [24], such large physical dimensions remain a major obstacle to practical integration within MRI systems [25], [26].

To overcome these challenges while maintaining the beneficial electromagnetic properties of dielectric structures, recent studies have explored metasurfaces (MSs) [8], [11], [12], [27], [28], [29] a class of metamaterials, which are thin artificial structures consisting of periodically arranged (typically metal) meta-atoms with electrically-small size and period, serving as much more compact and lightweight alternatives. Metasurfaces are widely used in different applications in microwave and antenna systems, for example, for miniaturization, gain enhancement [30], and wireless communication [31]. The transition from planar to cylindrical metasurfaces has been an important research direction in antenna engineering [32], [33], as cylindrical conformal structures enable better integration with compact equipment and improved control over wave propagation. This progression naturally extends to MRI applications, where a cylindrical shape is essential for focusing RF fields inside the body.

In this work, we proposed a novel lightweight and compact cylindrical MS acting as an artificial slow-wave structure, the design is inspired by the principles of compact, lightweight, and conformal artificial dielectric pad [35]. The cylindrical MS comprises a grid of copper strips loaded with parallel-plate capacitors and can replace the dielectric waveguide in the task of concentrating the $B_1^+$ field produced in the single-channel TW regime at 7 T in the human brain. By carefully engineering the MS properties, we were able to reach good equivalence between the MS and the DW. This approach preserves the advantages of dielectric-based TW MRI and improves practical feasibility for ultra-high-field human brain imaging. A preliminary version of this work was reported [36].

## II. Methods

### A. Numerical optimization of metasurface unit cell

Electromagnetic (EM) simulations were performed using the finite-integration-technique in the time domain (FIT-TD), as implemented in CST Studio Suite 2024 (Dassault Systèmes, Vélizy-Villacoublay, France). The cylindrical MS was designed to emulate the electromagnetic behavior of a high-permittivity DW previously reported in [20], provided that both structures have the same length $L$ = 170 mm in the axial direction. The reference DW [20] consisted of rectangular dielectric slabs with length $L_{DW}$ = 348 mm and thickness $W$ = 16 mm was further optimized for both practical fabrication and electromagnetic performance. Specifically, the relative permittivity was changed to $\varepsilon_r$ = 78, which is close to that of deionized water and therefore enables more convenient experimental implementation. In addition, the geometry was redesigned to use 8 slabs with an adjusted inner diameter $D$ = 270 mm to maintain better rotational symmetry, which is more suitable for supporting circularly polarized wave propagation. The MS unit cells were designed as crossed orthogonal copper strips (with a width of 0.3 mm) loaded with capacitors with a lumped capacitance of $C$, as proposed in [35]. To evaluate the equivalence between the MS and the DW, sections of both structures were simulated when placed inside a parallel-plate waveguide of the length $L$ (the same method as discussed in [35]). In the first configuration, a section of dielectric material of thickness $W$ was placed between two metal plates of the waveguide made of perfect electric conductor (PEC) as shown in Fig. 1(a). In the second configuration, a single row of $N$ = 5 MS unit cells was arranged along the longitudinal ($z$) direction of the same waveguide such that the vertical strips are connected to the waveguide plates (Fig. 1(b)). Note that the distance between the plates oriented in parallel to the $xz$-plane in both configurations was equal to the unit-cell period $T$ = 34 mm.

The period was chosen to provide that the gap between the cylindrical MS and the subject is larger than $T$ (to avoid periodic small-scale inhomogeneity of $B_1^+$ in the subject) and the length $L$ is an integer multiple of the unit cell. The $y$-directed strips in each unit cell are parallel to the electric field polarization direction in the waveguide. As discussed in [35], to keep the effective periodicity of the MS in the $y$ direction while simulating just one row between two PEC planes, two series-connected capacitors with double the capacitance (2$C$) are simulated instead of a single capacitor with capacitance $C$.

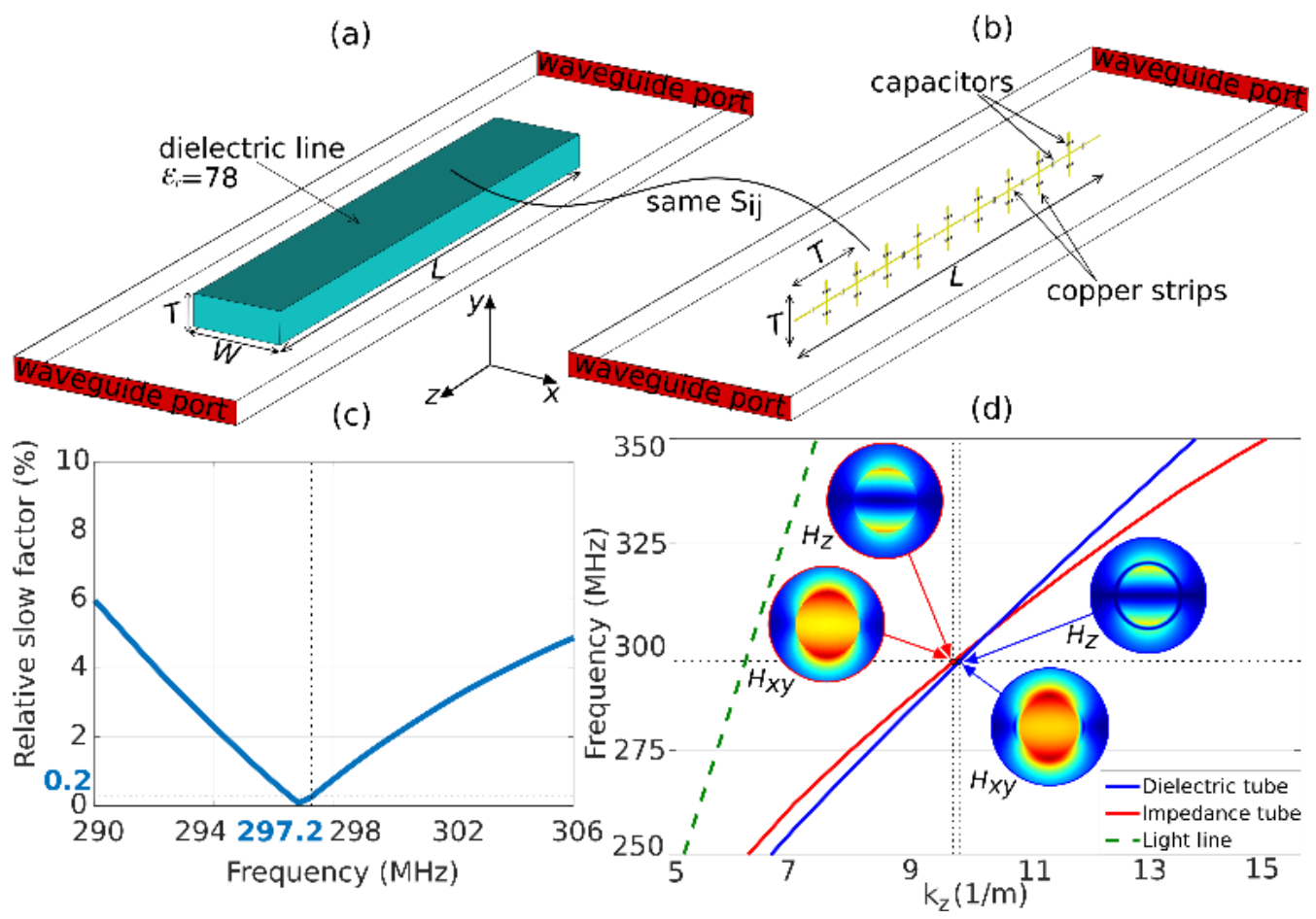

**Fig. 1.** (a) Numerical model of a dielectric slab inside a parallel-plate waveguide in CST Studio, (b) numerical model of a single row of MS' unit cells placed in the same parallel-plate waveguide, (c) numerically calculated relative slow factor (in percent) as a function of frequency, (d) dispersion (relative to the light line) and normalized $H$-field maps for the hollow impedance tube (mimicking the cylindrical MS) and a hollow dielectric tube [20] (mimicking the DW).

The equivalence between the MS and the DW was evaluated by comparing the phase shift for a wave traveling in the $z$-direction between two waveguide ports at the Larmor frequency $f$ = 297.2 MHz. The MS unit cell parameters, i.e. the period ($T$) and capacitance ($C$) were adjusted to match the DW phase shift. This comparison ensures only an equivalence for a wave propagating along the waveguide in the $z$-direction with a $y$-polarized electric field and an $x$-polarized magnetic field. However, due to in-plane isotropy of the MS and the considered dielectric slab, this equivalence holds for every propagation direction parallel to the $yz$-plane plane under the same polarization conditions.

The phase shift was extracted from the transmission coefficient as $\varphi = \arg(S_{12})$ at $f$ = 297.2 MHz. To quantify the difference in the slow-wave factors of the two structures, we introduced the relative slow factor (RSF), defined as RSF = $(\varphi_{DW}-\varphi_{MS})/\varphi_{DW}$. The optimal capacitance $C$, resulting in the best correspondence between the MS and DW, is achieved when RSF = 0. Fig. 1(c) shows the RSF as a function of frequency. It proves that the optimized MS with $C$ = 3.9 pF provides the same phase shift as the reference DW with an accuracy of 0.2% at 297.2 MHz.

### B. Physical principles

The dispersion analysis was performed using the Eigenmode Solver in CST Studio Suite 2024 to confirm and explain the equivalence between the MS and DW in the cylindrical configuration (in addition to the equivalence between the corresponding flat structures, as discussed in the previous subsection). The cylindrical MS was considered in the model as an infinitely long cylindrical surface of radius 140 mm with a grid impedance $Z_g$. The frequency-dependent value of this impedance was obtained for the corresponding flat MS as discussed below and attached to the cylindrical surface (transparent-type impedance boundary condition in CST). $Z_g$ equal to 0.46-i·118.3 Ohm at the Larmor frequency was extracted from the numerically calculated reflection and transmission coefficients of a normally incident plane wave [37], where unit cell surrounded with four periodic boundary conditions was excited using two Floquet ports. As a reference structure, an infinitely long hollow dielectric tube [20] with a wall thickness of 28 mm, an inner diameter of 250 mm and a relative permittivity of 21 was used to represent the DW. Both structures support a similar type of slow propagating waves characterized with a uniform magnetic field profile near the axis of the cylinder. The excitation of that fundamental mode explains the effect of $B_1^+$ concentration in the ROI when both structures (with finite length) are placed in the MRI bore. Therefore, to confirm that the cylindrical MS is equivalent to the cylindrical DW, one should compare the dispersion in both structures and the mode field profiles supported.

The calculated dispersion curves (Fig. 1(d)) demonstrate the behavior of the longitudinal wavenumbers $k_z$ for the MS and the DW. Near the Larmor frequency, the longitudinal wavenumbers exhibit close agreement. Specifically, at 297.2 MHz, the values reach approximately 9.6 m$^{-1}$, which is 54% higher than the free-space wavenumber (dashed green line in Fig. 1(d)), confirming that both structures support propagation with practically the same slow-wave factor. The equivalence between the MS and the DW is further confirmed by the comparison of their magnetic field (*H*-field) mode distributions. At the operating frequency of 297.2 MHz, the *H*-field maps (see the inset of Fig. 1(d)) reveal that both structures support the hybrid $HE_{11}$ mode with almost identical field profiles of similar field components.

Given that both the longitudinal wavenumbers and the field profiles exhibit such a high degree of correlation, it can be concluded that the electromagnetic properties of the structures remain consistent despite the minor change in geometry and finite length further used in the MRI bore environment.

### *C. Construction of the cylindrical metasurface*

After defining the desired unit cell parameters, a full numerical model of the cylindrical MS mounted on a hollow polycarbonate cylindrical holder ($\varepsilon = 2.9$, $\mathrm{tg}\delta = 0.01$) with a wall thickness of 5 mm and an inner diameter of 270 mm was constructed (Fig. 2(a)). The reference DW previously reported in [20] is shown in Fig. 2(b).

Simulations were performed using the Ella multi-tissue voxel model with material properties defined at 300 MHz [38] (Zurich MedTech, Zurich, Switzerland). The Ella voxel model was employed in the simulations to match the head size of the subjects included in the experimental studies in the presence of the proposed MS. Although minor anatomical differences exist among head models, the resulting trends in $B_1^+$ distribution are expected to be consistent.

A stepped-diameter circular waveguide was employed to model a realistic MRI-embedded waveguide, comprising an outer copper cylindrical shield corresponding to the cryostat

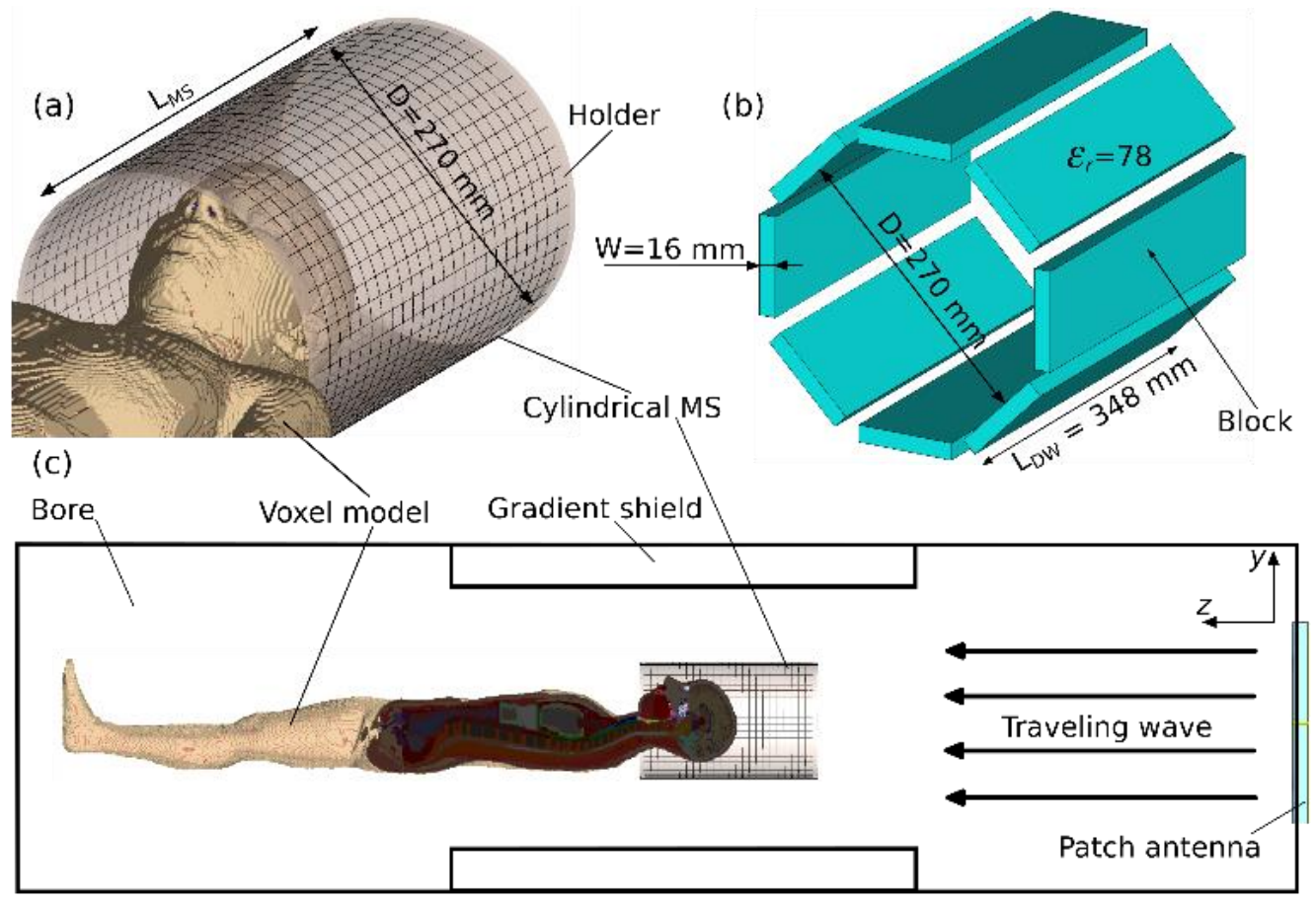

**Fig. 2.** (a) The general view of the cylindrical metasurface, (b) the reference DW previously reported in [15]; (c) the general view of the entire numerical model, including the cylindrical MS, Ella voxel model, and the embedded circular waveguide.

(900 mm in diameter and 3360 mm in length) and a concentric inner copper shield representing the RF shield (685 mm in diameter and 1220 mm in length). A circularly polarized (CP) patch antenna was used to excite a CP-polarized traveling wave. Fig. 2(c) presents the complete numerical model.

### *D. Optimization of the cylindrical metasurface*

The cylindrical MS (Fig. 2(a)) consisting of unit cells optimized as discussed in Subsection II-A is not fully equivalent to the DW previously reported in [20] (Fig. 2(b)) due to fundamental structural differences. The DW is discrete, comprising multiple flat dielectric blocks of thickness $W = 16$ mm and relative dielectric permittivity $\varepsilon_r = 78$ separated by air gaps. In contrast, the proposed MS is a uniform periodic structure, in which the optimized unit cells with lumped capacitors $C = 3.9$ pF occupy the cylindrical surface of diameter $D = 270$ mm. This difference affects field propagation. Consequently, the MS geometry must be parametrically tuned to achieve its desired performance in the configuration shown in Fig. 2(c).

Specifically, the cylindrical MS was tuned by its length $L_{MS}$ in the *z* direction and a longitudinal shift $\delta z$ relative to the head model until reaching the best compromise between the Tx-efficiency and RF field homogeneity in the brain region of the Ella voxel model. During optimization, $L_{MS}$ was varied across the following values: 340 mm, 410 mm, and 480 mm. The longitudinal shift $\delta z$ of the MS relative to the Ella head model was also varied, taking on the following values: 0 mm, 20 mm, 40 mm, and 60 mm; $\delta z = 0$ mm corresponded to the edge of the MS, located opposite the crown, aligning with the chin along the *z*-axis. By varying the $L_{MS}$ and $\delta z$, the RF field distribution within the brain region and level for a given excitation power at the input of the patch antenna can be controlled.

For each configuration of MS, the RF field ($B_1^+$) was calculated to assess both homogeneity and Tx-efficiency within voxel model brain tissues. Homogeneity was quantified using the coefficient of variation (COV), defined as the ratio of the standard deviation (STD) to the mean of $B_1^+$ value. Tx-efficiency was calculated as the average $B_1^+$ value over the ROI in brain tissues in the 180-mm region. Safety was assessed by calculating the SAR using the CST Legacy 10g-averaging method. Additionally, SAR-efficiency was defined as the ratio of the mean $B_1^+$ field to the square root of peak SAR. $B_1^+$ and $SAR_{10g}$ distributions were normalized to 1 W of power accepted by the system.

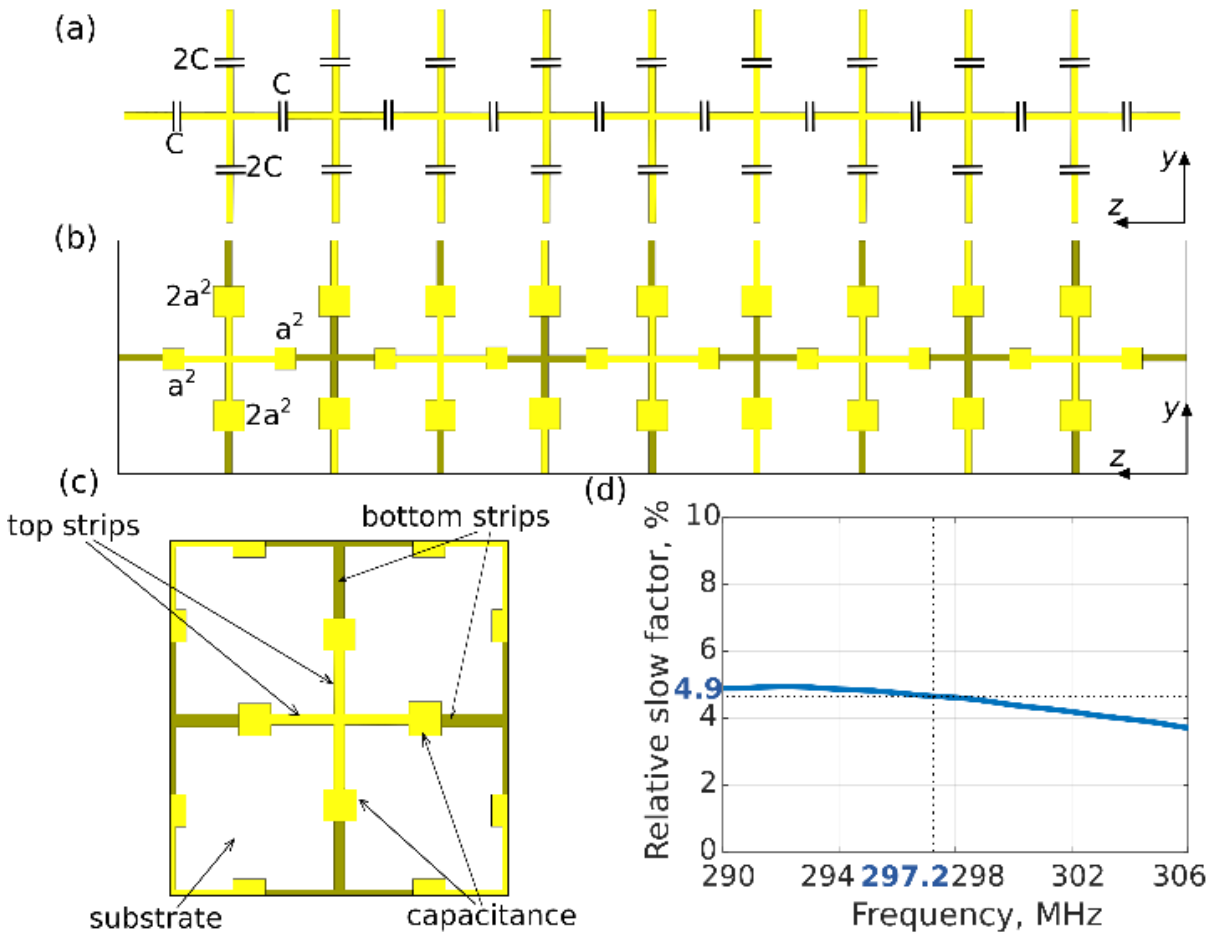

**Fig. 3.** (a) The numerical model of a single row of MS unit cells with lumped capacitors between two parallel metal plates, (b) the experimental PCB topology with copper strips loaded with parallel-plate capacitors, (c) the experimental realization of the unit cell of MS based on a dual-sided printed circuit board, (d) numerically calculated the relative slow factor between the lumped-element and PCB realizations (in percent) as a function of frequency.

For comparison with the proposed MS structure, the reference DW configuration (the same as proposed in [20]) was also simulated under the same excitation conditions. The proposed MS or the reference DW are designed to increase the efficiency of RF energy delivery. As shown in [20], an important aspect of this enhancement is the effective mode conversion, which involves maximizing the transverse component of the magnetic field ($H_{xy}$). Consequently, the ratio $H_{xy}/|H|$ is maximized in the region of interest. The impact of the proposed MS on this transverse magnetic field enhancement is observes to be similar to that of the DW.

### *E. Experimental realization of the metasurface*

To simplify fabrication, reduce costs, and improve design robustness, the technology of flexible printed-circuit boards (PCBs) was employed. In particular, all strips of the MS unit cells were printed on flexible polyamide film (thickness 0.05 mm, $\varepsilon = 3.5$, tg$\delta$ = 0.0027). The lumped capacitors were replaced with parallel-plate printed capacitors, which plates are located on the opposite sides of the polyamide film. Fig. 3(a) shows the model of a row of MS unit cells (the same as discussed in Subsection II-A) with period $T = 34$ mm and with lumped capacitors, placed within a parallel-plate waveguide. As mentioned previously, two series-connected capacitors with a capacity of $2C$ each in the $y$-direction are modeled, effectively modeling an isotropic MS in the presence of the walls of the parallel-plate waveguide.

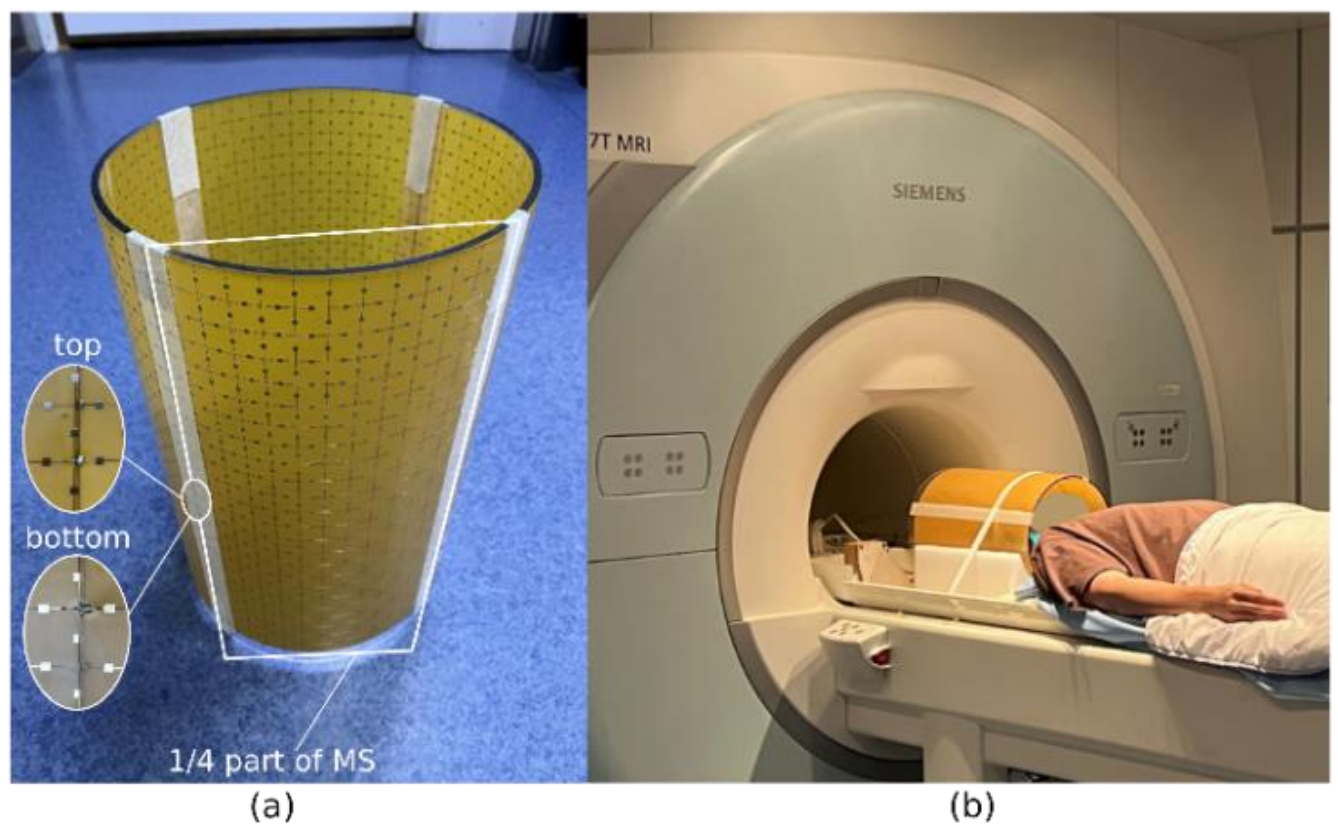

**Fig. 4.** (a) A prototype of the cylindrical metasurface based on four connected PCB segments mounted on a polycarbonate holder, (b) the experimental setup for *in vivo* imaging.

For the experimental realization of the MS, copper strips (0.035-mm-thick metallization) were patterned on both sides of the polyamide film. The parallel-plate capacitors consisted of square copper plates (side length $a$) printed on both sides of the same film and connected to the copper strips, which are also use both top and bottom sides of metallization (see Fig. 3(b) and 3(c)). A capacitance of $C = 3.9$ pF, equivalent to that of the optimal lumped element value, was achieved with a side length of $a = 2.4$ mm for the parallel-plate capacitors, based on the same slow factor comparison in a parallel-plate waveguide as discussed in Subsection II-A at $f = 297.2$ MHz. Note that the parallel-plate capacitors in the strips oriented in the $y$-direction effectively had twice the copper plate area ($2a^2$) compared to the capacitors inserted to the strips oriented in the z-direction ($a^2$). This is done to model an effectively isotropic MS in the presence of the walls of the parallel-plate waveguide. Note that for the cylindrical MS all capacitors have the same area ($a^2$) with a side length of 2.4 mm. The practical realization of the unit cell is shown in Fig. 3(b).

The frequency dependence of the RSF (Fig. 3(d)) demonstrates high similarity between the MS realizations with lumped capacitors ($C = 3.9$ pF) and a parallel-plate one with $a = 2.4$ mm in terms of phase velocity.

### *F. The prototype of the cylindrical metasurface*

The experimental realization of the proposed MS was constructed. A photograph of the experimental sample is shown in Fig. 4(a). The geometry and dimensions of the cylindrical polycarbonate holder, size of the parallel-plate capacitors, and the width of copper stripes, as well as the properties of the polyamide substrate were identical to those used in the numerical simulations (see Subsection II-E). The cylindrical MS consisted of four identical flat PCBs, which were connected together in a uniform cylindrical periodic structure mounted on the outer surface of the holder by soldering the strips terminations. For proper connection between the separate PCBs, each PCB had metalized vias to provide electrical connectivity between the top and bottom layers near the corresponding connected edges (see the inset of Fig. 4(a)). This suggests that the PCB prototype closely approximates the numerical model of the proposed cylindrical MS.

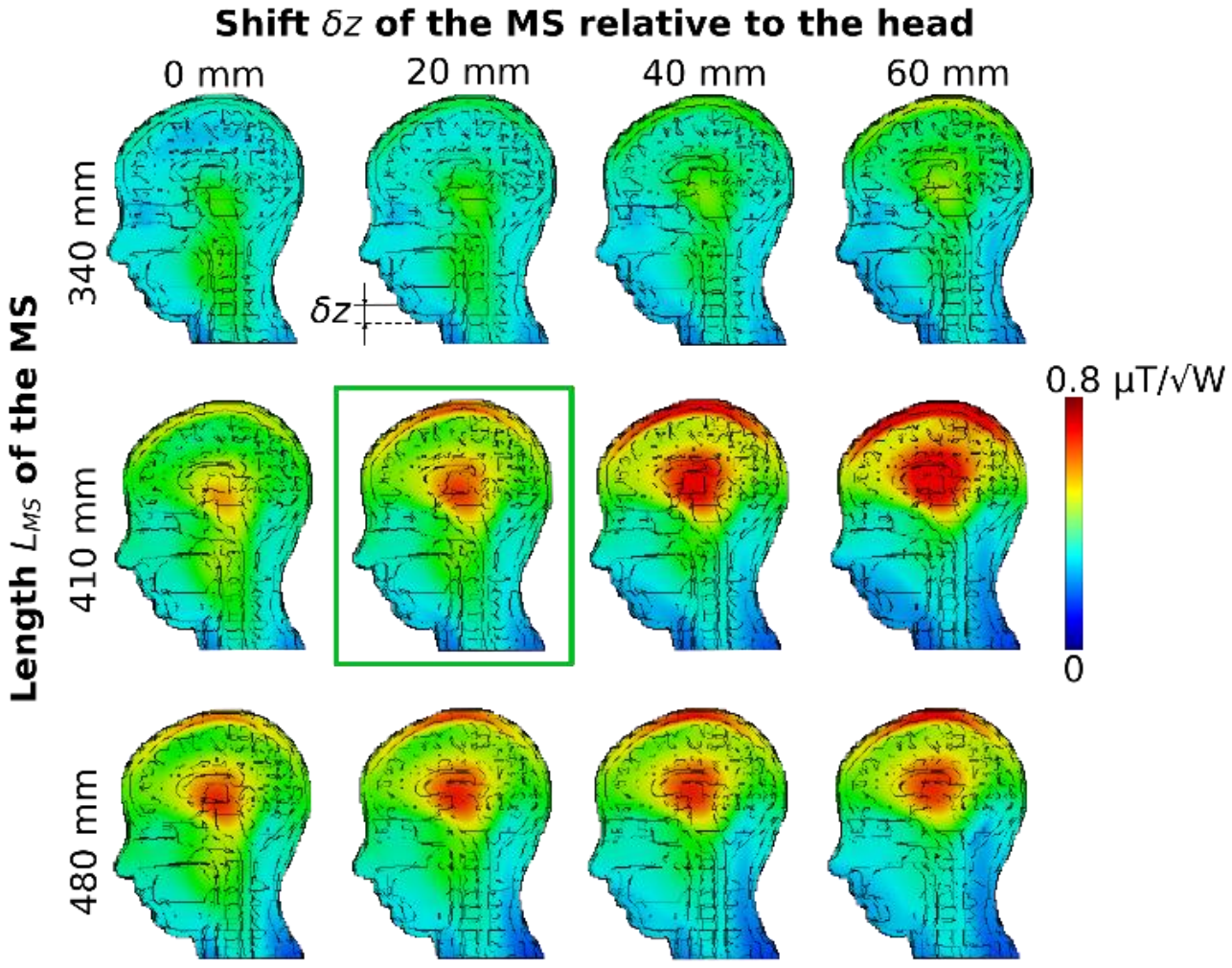

**Fig. 5.** Numerically simulated $B_1^+$ maps in the central sagittal slice of the Ella voxel model using different lengths and the longitudinal shift of the cylindrical metasurface relative to the head.

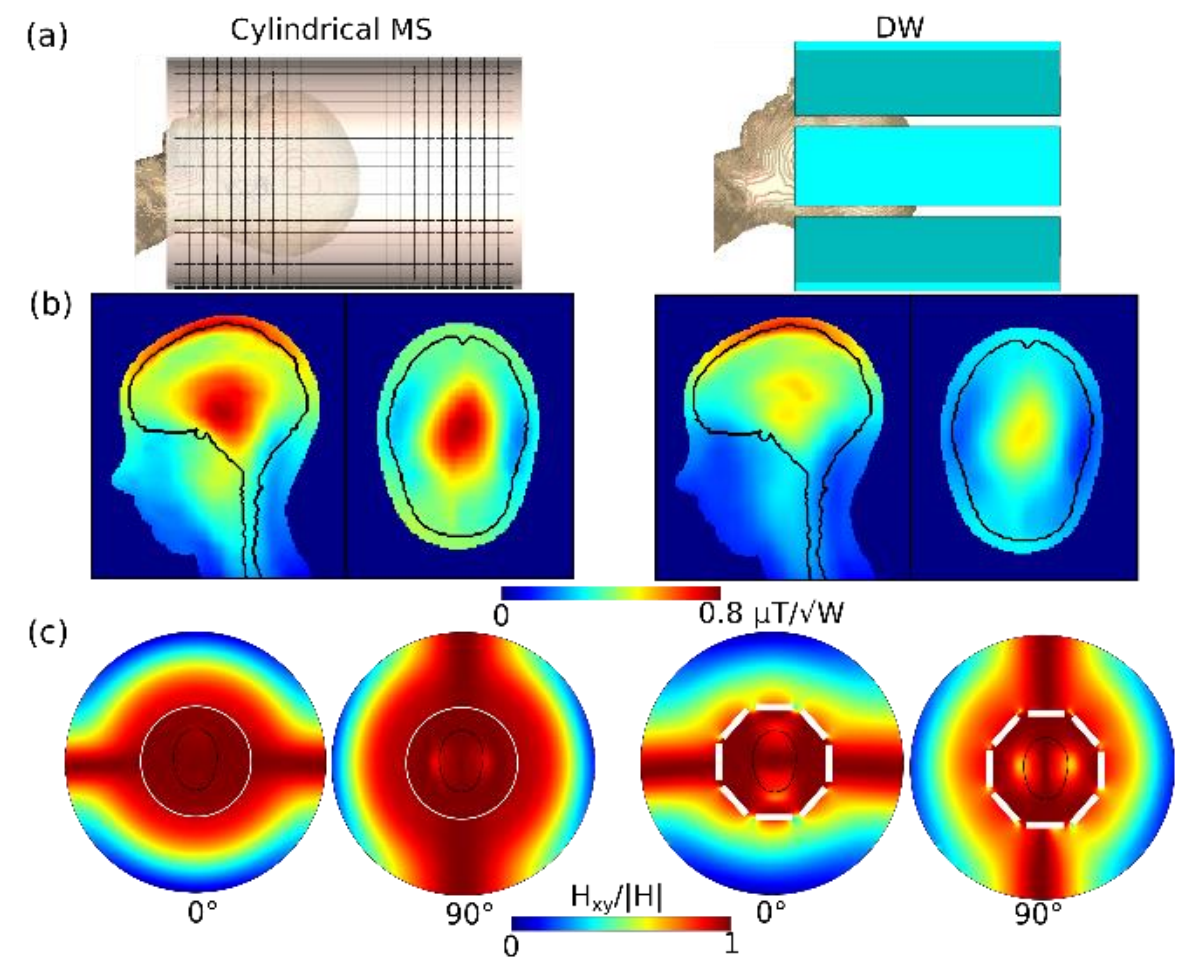

**Fig. 6.** (a) Numerical models in CST Studio of the proposed MS ($L_{MS}$ = 410 mm) and the dielectric waveguide with Ella voxel model, (b) central sagittal and axial $B_1^+$ maps obtained using simulated MS and dielectric waveguide, (c) the ratio of transverse and the longitudinal magnetic field components in MS and in a dielectric waveguide inserted into the circular metallic waveguide.

### G. *Phantom and in vivo measurements*

MRI experiments were performed on a 7 T human MRI scanner (Siemens MAGNETOM 7 T, Siemens Healthineers, Erlangen, Germany) operating in clinical mode using single-channel transmission, consistent with the methodology from [20]. A brain-tissue-mimicking anthropomorphic head phantom [39] was used for initial imaging studies. Subsequently, *in vivo* imaging was conducted on a healthy volunteer (24 y.o., 180 cm, 85 kg, male), with all procedures approved by the Ethics Committee at Zhejiang University (2022-45). Written informed consent was obtained from the subject prior to the study.

For radiofrequency excitation, a circular patch antenna was used to deliver electromagnetic waves through the MRI bore circular waveguide. The antenna, positioned at the service end of the bore, consisted of a circular copper patch (350 mm in diameter) and a ground sheet interleaved with two acrylic slabs. The patch antenna was positioned at the same distance as in the numerical model. Operating in transceiver mode via a transmit/receive (T/R) switch, the patch antenna was driven by two feeding ports interfaced with a 90° quadrature hybrid to provide CP-mode excitation of TW. Both the cylindrical MS (see Fig. 4(b)) and, for comparison, the reference DW [20] were placed on the patient table and positioned at the isocenter of the MRI magnet during the scan.

Quantitative $B_1^+$ mapping using an actual-flip-angle (AFI) sequence [40] ($TR_1$/$TR_2$: 20 ms/50 ms; *TE*: 2.5 ms; voxel size: 1.8 mm × 1.8 mm × 3 mm) was employed to evaluate excitation efficiency measure of the $B_1^+$ field. In the *in vivo* study, GRE T2* images (*TR*: 450 ms, *TE*: 5 ms, nominal flip angle: 60°; voxel size: 1.5 mm × 1.5 mm × 2 mm) were acquired to qualitatively assess excitation homogeneity.

## III. Results

### A. *Optimization of the cylindrical metasurface*

Fig. 5 presents the numerically calculated $B_1^+$ distributions in the central sagittal plane of the Ella voxel model for different combinations of $L_{MS}$ and $\delta z$ in the longitudinal direction.

These parameter sets were selected to illustrate the variation of the field distribution within the explored parameter range. The configuration providing the best balance between homogeneity and Tx-efficiency ($L_{MS}$ = 410 mm and $\delta z$ = 20 mm) is highlighted by a green contour. Reported quantities include $B_1^+$ homogeneity ($B_1^+$ COV), Tx-efficiency ($<B_1^+>$), $pSAR_{10g}$, and SAR-efficiency. The mean values and standard deviations were calculated specifically for brain tissues.

Table I summarizes the calculated performance metrics for the proposed cylindrical MS with the investigated parameters using the Ella voxel model.

For the numerical models partially shown in Fig. 6(a), the numerically calculated $B_1^+$ distributions for the optimal cylindrical MS and the reference DW [20] are compared in Fig. 6(b). The figure also shows the ratio of transverse to longitudinal magnetic field components within the circular metallic bore waveguide for the Ella voxel model (two linear polarization modes of excitation, i.e., 0° and 90°, are shown separately for clarity). The maps of the ratio $H_{xy}/|H|$ (Figure 6(c)) were obtained using MATLAB 2022 and data exported from CST Studio.

A quantitative comparison of the calculated field quantities for both structures is presented in Table II. As shown, the proposed MS outperforms the reference DW structure.

TABLE I
OPTIMIZATION RESULTS OF THE CYLINDRICAL METASURFACE LOADED BY THE ELLA VOXEL MODEL FOR VARIOUS LENGTHS AND LONGITUDINAL SHIFTS OF THE CYLINDRICAL METASURFACE RELATIVE TO THE HEAD

| $L_{MS}$, mm | δz, mm | $B_1^+$ COV | $< B_1^+ >$, μT | $pSAR_{10g}$, W/kg | SAR-eff., μT/√W/kg |
|---|---|---|---|---|---|
| 340 | 0 | 0.228 | 0.246 | 0.416 | 0.381 |
| | 20 | 0.225 | 0.244 | 0.346 | 0.415 |
| | 40 | 0.222 | 0.274 | 0.298 | 0.502 |
| | 60 | 0.245 | 0.305 | 0.278 | 0.578 |
| **410** | 0 | 0.231 | 0.370 | 0.444 | 0.555 |
| | **20** | **0.249** | **0.432** | **0.531** | **0.593** |
| | 40 | 0.281 | 0.479 | 0.779 | 0.543 |
| | 60 | 0.322 | 0.505 | 1.013 | 0.502 |
| 480 | 0 | 0.257 | 0.422 | 0.512 | 0.590 |
| | 20 | 0.277 | 0.431 | 0.599 | 0.557 |
| | 40 | 0.301 | 0.430 | 0.678 | 0.522 |
| | 60 | 0.331 | 0.425 | 0.738 | 0.495 |

TABLE II
COMPARISON BETWEEN THE CYLINDRICAL METASURFACE AND THE DIELECTRIC WAVEGUIDE

| Type | $B_1^+$ COV | $< B_1^+ >$, μT | $pSAR_{10g}$, W/kg | SAR-eff., μT/√W/kg |
|---|---|---|---|---|
| MS | 0.249 | 0.432 | 0.531 | 0.593 |
| DW | 0.301 | 0.339 | 0.495 | 0.482 |

### *B. Measurements*

The performance of the proposed MS in the TW-setup was evaluated using an anthropomorphic head phantom. $B_1^+$ maps were derived for the conventional TW excitation setup with the patch antenna (without any passive structures), with the fabricated cylindrical MS and with the reference DW [20]. The resultingimages are shown in Fig. 7.

After the phantom studies, *in vivo* experiments were performed on a healthy volunteer, replicating the same cases. We acquired GRE images (Fig. 8) to qualitatively analyze the resulting effect of the RF field distribution patterns to image quality.

## IV. DISCUSSION

The proposed cylindrical MS for the TW excitation setup at 7 T has unit cells realized using flexible PCBs with parallel-plate capacitors similar to the flat MS for improving homogeneity of abdominal imaging at 3 T [35] (with a birdcage excitation). Similarly, the equivalence in terms of slow-wave factor is required to replace an original all-dielectric structure with a light and flexible MS. This equivalence can be ensured using a simplified simulation approach in a parallel-plate waveguide configuration (the same as used in [35]). However, numerical simulation in the TW MRI configuration with a voxel model demonstrated that the longitudinal extent and shift $\delta z$ of the MS relative to the head affect both the distribution shape and level of the $B_1^+$ field within the brain. Therefore, both parameters were tuned (for Ella voxel model) while keeping the previously chosen microstructure of the MS to obtain the best balance between Tx-efficiency and field homogeneity.

Our parametric analysis reveals a clear trade-off between these two metrics. Shorter MS configurations tend to promote improved $B_1^+$ homogeneity, whereas longer structures enhance Tx-efficiency by increasing effective RF power delivery into the brain region. However, excessive longitudinal extension can lead to less uniform field distributions. The MS configuration with a length $L_{MS}$ = 410 mm and a longitudinal shift $\delta z$ = 20 mm was identified as an optimal compromise, providing a balanced improvement in both Tx-efficiency and $B_1^+$ homogeneity.

A comparison with the reference DW (which was previously optimized using the same metrics [20]) emphasized the advantages of the proposed cylindrical MS. The cylindrical MS demonstrated a simultaneous improvement in terms of homogeneity, Tx-efficiency, and SAR-efficiency. In addition, the MS avoids water-related artifacts, as it creates no MR signal. Finally, using the MS makes the setup more comfortable for the patient, since it has considerable smaller thickness (even in the presence of the polycarbonate holder) than for the DW.

The evaluation of field homogeneity, as measured by the $B_1^+$ COV, improved by 17.3% (decreasing from 0.301 to 0.249) compared to the DW in the Ella voxel model. Furthermore, the Tx-efficiency ($<B_1^+>/\sqrt{P_{acc}}$) of the proposed MS is 27.4% higher than that of the DW (0.432 μT and 0.339 μT, respectively). While numerical simulations indicated that the MS produced a higher $pSAR_{10g}$ than the DW (by 7.3%, from 0.495 W/kg to 0.531 W/kg), the proportional increase in Tx-efficiency for the MS, relative to the DW, was greater than the corresponding increase in $pSAR_{10g}$. As a result, the SAR-efficiency of the proposed MS loaded by the Ella voxel model was 23% higher (0.593 μT/√W/kg compared to 0.482 μT/√W/kg) than that calculated for the DW.

Phantom $B_1^+$ maps acquired showed that the incorporation of the cylindrical MS leads to a pronounced improvement in RF field homogeneity and Tx-efficiency compared to the conventional TW excitation without passive structures, which is in close agreement with numerical predictions.

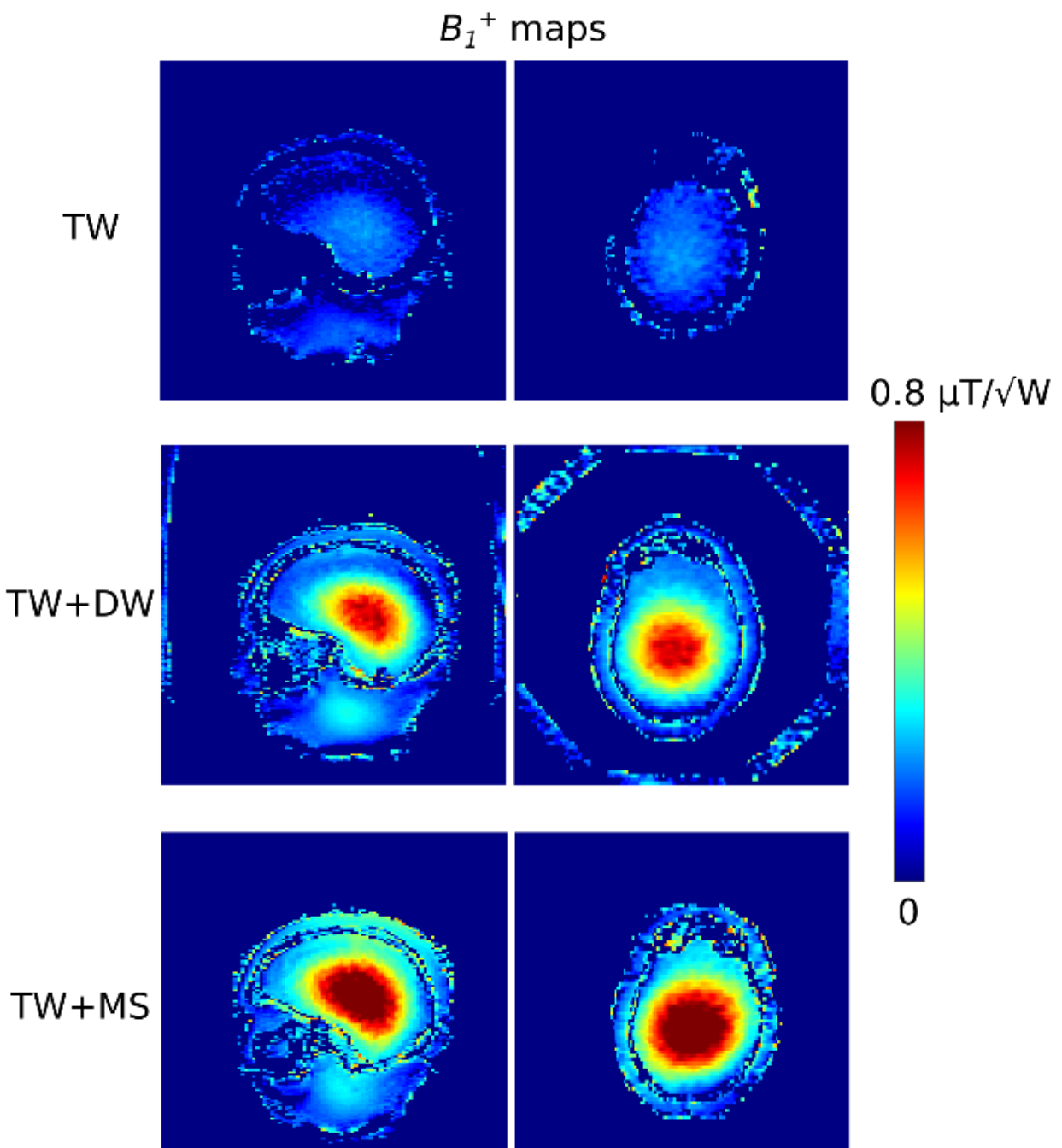

**Fig. 7.** MRI experimental results of quantitative $B_1^+$ maps acquired from an anthropomorphic head phantom in the central sagittal and axial planes.

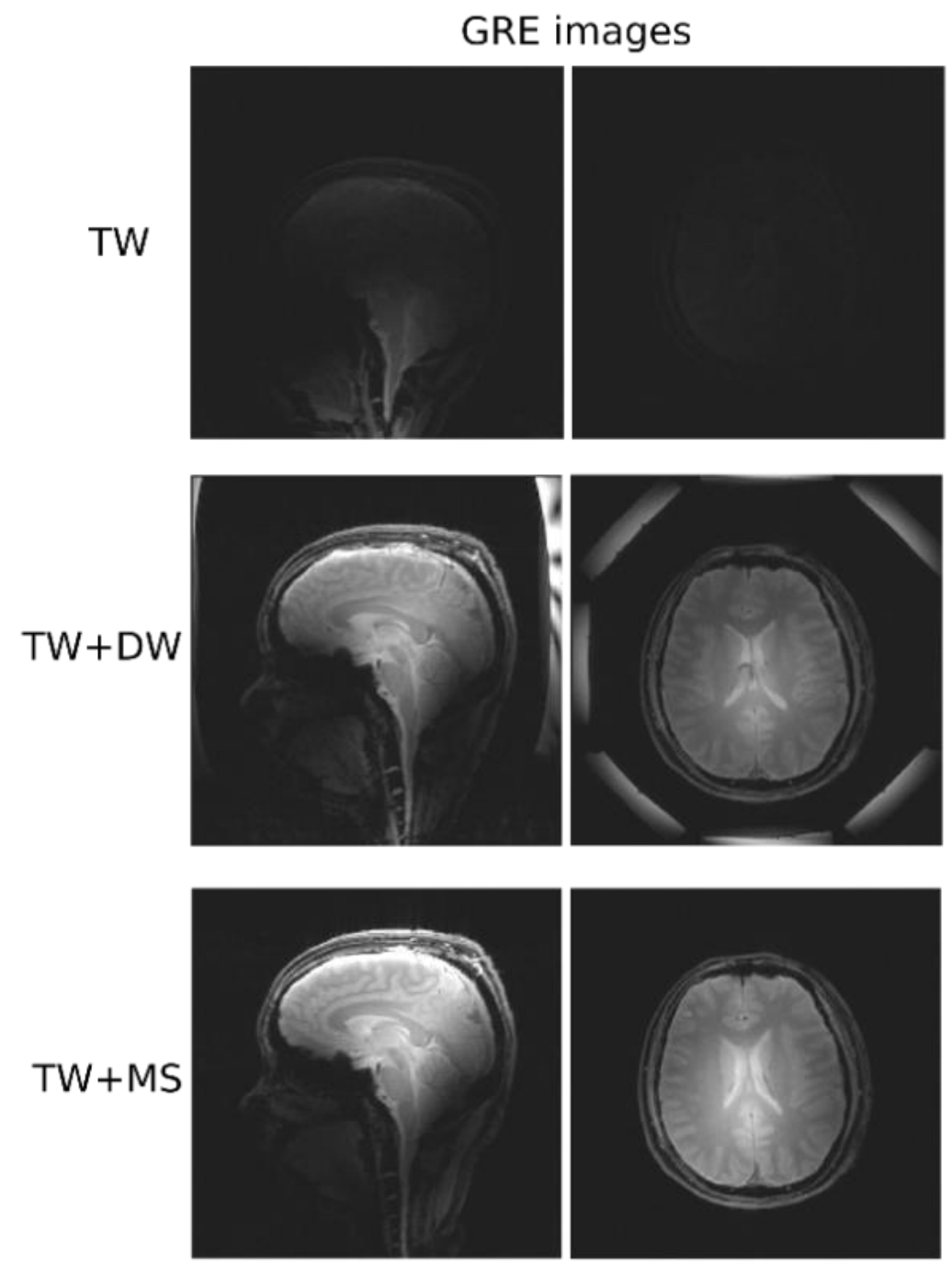

**Fig. 8.** MR-images using the proton-density weighted GRE sequence acquired from a human subject *in vivo* in the central sagittal and axial planes.

*In vivo* experiments proved the feasibility of the proposed approach. GRE images acquired on the *in vivo* subject with the MS exhibit increased signal intensity compared with both the DW-based and conventional TW configurations, consistent with the enhanced Tx-efficiency observed in the phantom experiments. These findings experimentally confirm that the MS effectively reshapes and concentrates the $B_1^+$ field within the ROI.

## V. Conclusion

The optimized MS, consisting of a lightweight grid of copper strips loaded with parallel-plate capacitors, outperforms both classical TW MRI and the state-of-the-art DW in terms of $B_1^+$ homogeneity, Tx-efficiency, and SAR-efficiency. These improvements result in a more uniform and effective radiofrequency field distribution within human brain tissues.

Beyond its electromagnetic performance, the proposed MS offers important practical advantages. Its compact and lightweight structure reduces system bulk and eliminates water-related signal artifacts commonly associated with dielectric waveguides, thereby improving patient comfort and experimental robustness.

By enabling efficient RF excitation using a single transmit channel in clinical operating mode, the cylindrical MS provides a practical pathway toward improved image quality and broader applicability of traveling-wave MRI at ultra-high field strengths.

## Acknowledgment

The authors acknowledge the Zhejiang University 7 T Brain Imaging Research Center for technical support and thank Y. Lu, Y. Xie and B. Xu for their assistance with this work.

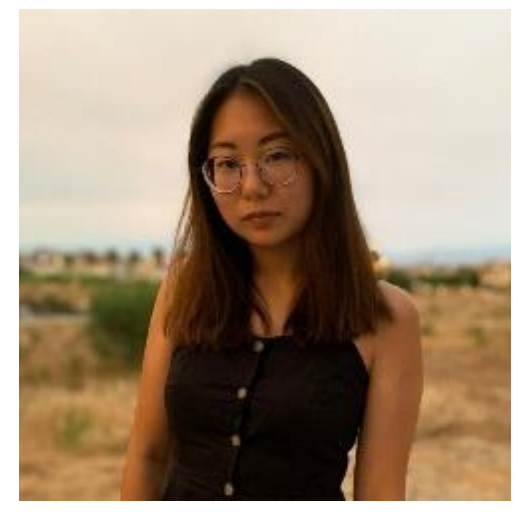

**Kristina I. Popova** was born in Saint Petersburg, Russia, in 2000. She received B.Sc. and M.Sc. degrees in technical physics from ITMO University, Saint Petersburg, Russia, in 2022 and 2024. Her research interests include the development radio frequency coils for ultrahigh-field magnetic resonance imaging.

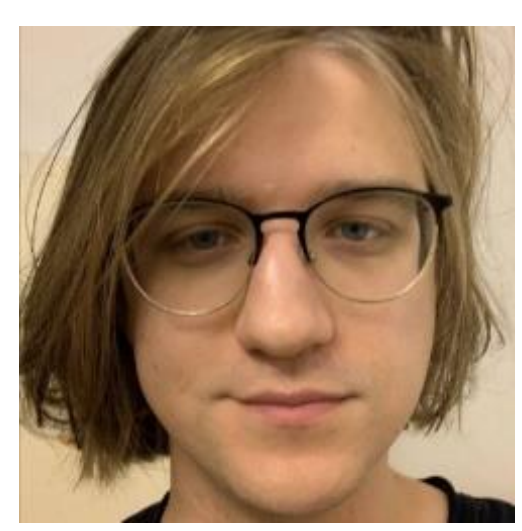

**Georgiy A. Solomakha** was born in Murmansk, Russia, in 1994. He received the M.Sc. degree in radiophysics from St. Petersburg Polytechnic State University, Saint Petersburg, Russia, in 2017, and the Ph.D. degree in antennas and microwave devices from the School of Physics and Engineering, ITMO University, Saint Petersburg, in 2021. He is currently a Post-doctoral researcher in Department of High-Field Magnetic Resonance, Max Planck Institute for Biological Cybernetics, Tübingen. His current research interests include magnetic resonance imaging (MRI) coils for ultrahigh-field MRI.

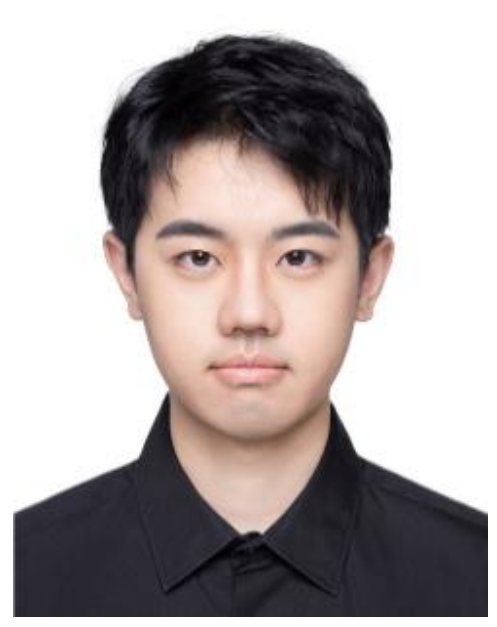

**Zicheng Wen** received the B.S. degree in electronic information engineering from Xidian University, Xi'an, China, in 2023. He is currently pursuing the M.S. degree in electronic and information technology at the Hangzhou Institute of Xidian University, Hangzhou, China. His current research interests include ultra-high-field magnetic resonance imaging, radiofrequency coil design, and artificial dielectric materials.

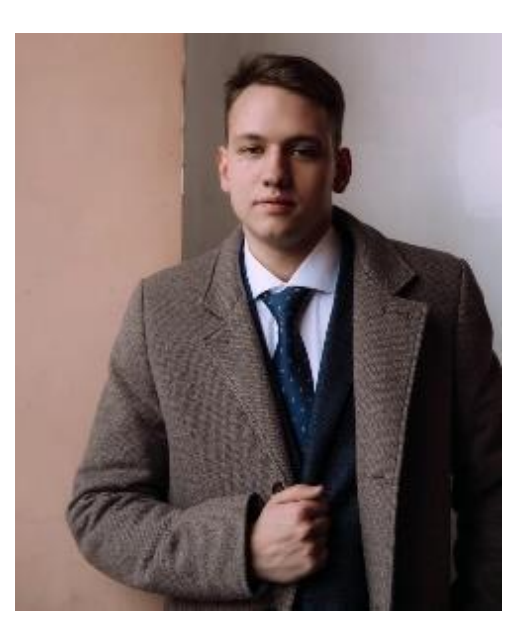

**Mikhail M. Popov** was born in Saint Petersburg, Russia, in 2001. He received B.Sc. and M.Sc. degrees in technical physics from ITMO University, Saint Petersburg, Russia, in 2022 and 2024. His research interests include the metasurfaces and digital communications.

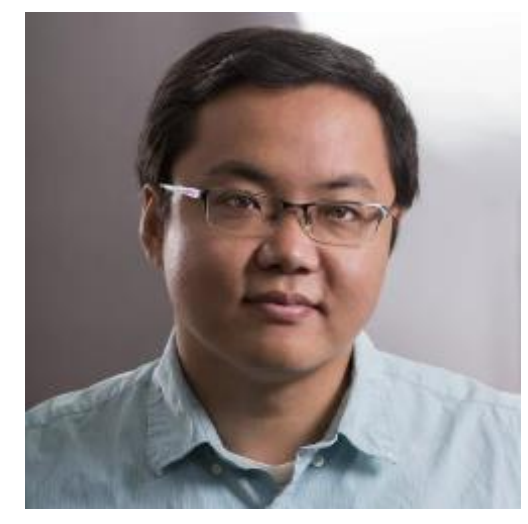

**Xiaotong Zhang** (Senior Member, IEEE) received the B.S. and Ph.D. degrees in Electrical Engineering from Zhejiang University, Hangzhou, China, in 2004 and 2009, respectively. From 2009 to 2015, he was a Postdoctoral Associate and later a Research Associate with the Department of Biomedical Engineering and the Center for Magnetic Resonance Research, University of Minnesota, Minneapolis, MN, USA. In 2015, he joined Zhejiang University, where he was promoted to tenured Associate Professor in 2024 and appointed Director of the Institute for Advanced Electrical and Electronic Technologies in 2025. Dr. Zhang currently serves as an Associate Editor for the IEEE Journal of Electromagnetics, RF and Microwaves in Medicine and Biology, and has been a member of the Editorial Advisory Board of NMR in Biomedicine since 2022. His research interests cover a wide range of areas, including biomedical imaging, electromagnetic modeling and analysis, radio-frequency electronics, and the development of artificial intelligence-based brain–computer interfaces.

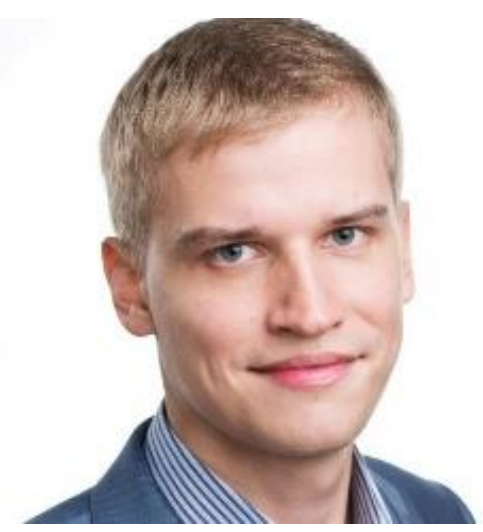

**Stanislav B. Glybovski** was born in Syktyvkar, Russia, in 1987. He received the M.Sc. and Ph.D. degrees in radiophysics from St. Petersburg Polytechnic State University, Saint Petersburg, Russia, in 2010 and 2013, respectively, and the D.Sc. (Habilitation) degree in radiophysics from the School of Physics and Engineering, ITMO University, Saint Petersburg, in 2023. He is currently a Leading Researcher with the School of Physics and Engineering, ITMO University. His current research interests include antennas and microwave devices, computational electromagnetics, metamaterials, and magnetic resonance imaging coils.

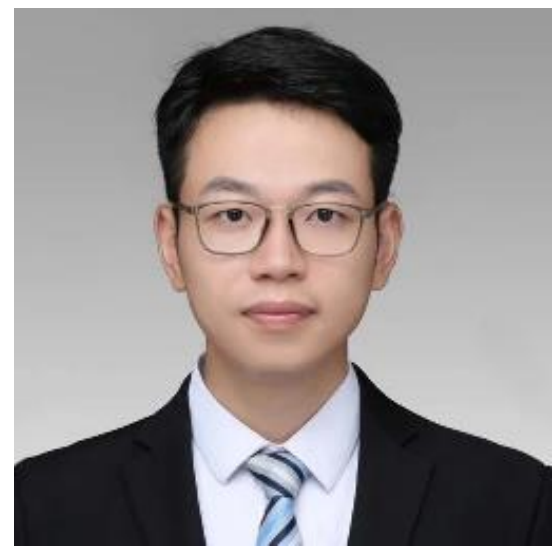

**Yang Gao** (Member, IEEE) received the Ph.D. degree in biomedical engineering from Zhejiang University, Hangzhou, China, in 2019. He has been involved in a two-year postdoctoral program in Anna Wang Roe's lab in Zhejiang University medical school. He is currently a tenure-track assistant professor in Hangzhou Institute of Technology, Xidian University, China. His research interests include electromagnetics in magnetic resonance imaging and its application in medicine.